\documentclass[a4paper,12pt]{article}
\usepackage[left=2.5cm,right=2.5cm,top=2.5cm,bottom=2.5cm]{geometry}
\begin{document}
\begin{center}
\textbf{Modified Ricci flow and asymptotically non-flat spaces}\\
\vspace{10mm}
Shubhayu Chatterjee$^{*}$ {\footnote{E-mail: shubhayu@iitk.ac.in}} 
and Narayan Banerjee$^{\dagger}$ {\footnote {E-mail: narayan@iiserkol.ac.in}}\\
$^{*}$ Department of Physics, Indian Institute of Technology, Kanpur;\\ Kanpur 208016; India.\\
$^{\dagger}$ IISER - Kolkata, Mohanpur Campus, P.O. BCKV Main Office, District Nadia,\\ West Bengal 741252, India.\\
\end{center}
\date{}
\vspace{10mm}
\begin{abstract}
The present work extends the application of a modified Ricci flow equation to an asymptotically non flat space, namely Marder's cylindrially symmetric space. It is found that the flow equation has a solution at least in a particular case.
\end{abstract}

\vspace{0.5cm}
\vspace{0.5cm}
\section*{Introduction:}
The Ricci Flow (RF) is an evolution equation for a Riemannian metric $g_{\mu \nu}$ with respect to a scalar parameter, say $\tau$, in terms of the Ricci tensors $R_{\mu \nu}$. The equation reads as
\begin{equation} 
\frac{\partial g_{\mu \nu}}{\partial \tau} = -2R_{\mu \nu}. 
\end{equation} 
RF equation, or some suitably modified form of it, finds a lot of application in gravitation theories\cite{6}. The Ricci flow equation has also been discussed in a cosmological scenario\cite{7}. Recently Ricci flow is used in black hole physics, particularly in the investigation of the stability of a black hole\cite{8}. In the context of Supergravity vacuum solutions, Ricci flow has been discussed by Hu et al\cite{9}. Ricci flow in the connection with a braneworld scenario has been investigated by Das etal\cite{10}. The reason for using the Ricci flow equation is the following.\\ 
For three-dimensional manifolds, if we expand the metric around a flat space given by the equation $g_{\mu \nu}=\eta_{\mu \nu}+h_{\mu \nu}$, we find that the equation (1), to linear order, yields
\begin{equation} 
\frac{\partial h_{\mu \nu}}{\partial \tau} = \nabla ^2 h_{\mu \nu}.
\end{equation}
This looks like a `heat conduction equation' for $h_{\mu \nu}$. This is indeed a parabolic equation and thus the solution tends to lose memory of the initial condition. It is well known in physics that a heat flow finally leads to a state with maximum entropy irrespective of the initial conditions. It is thus interesting to see if  the Ricci flow equation can also lead to a state of maximum entropy for the gravitating system given by $h_{\mu \nu}$, with the system forgetting the initial conditions when it reaches its final state. In order to bring out the correspondence between the statistical mechanis and gravity, Perelman proposed a modification of Ricci flow, known as the gradient formulation\cite{2}. The equation looks like 
\begin{equation}
\frac{\partial h_{\mu \nu}}{\partial \tau} = \nabla ^2 h_{\mu \nu} + 2D_{\mu}D_{\nu}f,
\end{equation}
where $f$ is a scalar on the manifold. However, as shown by Samuel and Roy Chowdhury \cite{1}, there exists no fixed point solution to these equations in the Schwarzschild space for any choice of fand thus does not give any extremum. They therefore concluded that Perelman's entropy function  is not connected to the Bekenstein-Hawking entropy for black holes. \\
They have also postulated a modification of Perelman's gradient formulation as 
\begin{equation} 
\frac{\partial g_{\mu \nu}}{\partial \tau} = -2fR_{\mu \nu}+2D_{\mu}D_{\nu}f 
\end{equation}
supplemented by an evolution equation for f, which we can relate to the scalar diffusivity in thermodynamics,
\begin{equation} 
\frac{\partial f}{\partial \tau}=D^2f.
\end{equation}
The only difference here from Perelman's formulation being the appearence of the function $f$ in the first term of the right hand side. The equations for the fixed points of the above flow are
\begin{equation} 
fR_{\mu \nu}=D_{\mu}D_{\nu}f ,
\end{equation}
\begin{equation} 
D^2f=0. 
\end{equation}
For the asymptotically flat exterior Schwarzschild space, the geometry for a static spherically symmetric mass distribution, the scalar function $f(r)=\left(1-\frac{2M}{r}\right)^\frac{1}{2}$ satisfies both the equations, where $M$ is the total mass of the spherical body. Thus the modified Ricci flow might have some connection with geometric entropy as its fixed points are extrema of the Bekenstein-Hawking entropy.\\
We try to extend this concept of a scalar diffusivity function to asymptotically non-flat spaces, using Marder's solution to Einstein's field equations for cylindrically symmetric space, and thereby to check if thermodynamical considerations could be dealt with for asymptotically non-flat spaces also.

\section*{The Scalar Diffusivity function:}
Marder \cite{3} gave the following exterior metric for a static cylindrically symmetric matter distribution
\begin{equation} 
ds^2= r^{2c}dt^2-B^2r^{2c(c-1)}\left(dr^2+dz^2\right)-r^{2(1-c)}d\phi^2. 
\end{equation}
Here $c$ is a parameter roughly twice the mass per unit length of the cylinder, and $B$ is another parameter related to $c$.\\
For some t = constant hypersurface, the spatial part of the metric is
\begin{equation} ds^2= B^2r^{2c(c-1)}\left(dr^2+dz^2\right)+r^{2(1-c)}d\phi^2 .\end{equation}
The corresponding non-zero Ricci tensors are
$$ R_{rr}=-\frac{c(c-1)^2}{r^2},$$
$$ R_{zz}=\frac{c^2(c-1)}{r^2},$$
$$ R_{\phi \phi}=-\frac{c(c-1)}{B^2}r^{-2c^2}.$$
We now want a solution to the equations (6) and (7). Consistent with cylindrical symmetry, we assume that f is a function of the radial coordinate $r$ only.\\
For $\mu =\nu =r$, we get 
\begin{equation} 
\frac{\partial^2f}{\partial r^2}-\frac{c(c-1)}{r}\frac{\partial f}{\partial r}+ \frac{c(c-1)^2}{r^2}f=0 .\end{equation}
For $\mu =\nu =z$ or $\phi$, we get
\begin{equation} 
\frac{\partial f}{\partial r}=\frac{c}{r}f .
\end{equation}
The equation for the covariant laplacian of f gives
\begin{equation} \frac{\partial}{\partial r}\left(r^{1-c}\frac{\partial f}{\partial r}\right)=0 .\end{equation}
Equation (12) readily integrates to yield (subject to a simple choice for the arbitrary constants of integration)
\begin{equation} f(r)=r^c , \end{equation} 
which satisfies both (10) and (11), and is hence a solution. Thus we have found a solitonic solution to the modified RF equations. The existence of such a scalar diffusivity indicates that we can pursue thermodynamics even for an asymptotically non-flat space. Equation (13) indicates that the scalar diffusivity function $f(r)$ attains the significance of the lapse function, as $f^{2}$ is equal to $g_{00}$. This is indeed consistent with the result obtained by Samuel and Roy Chowdhury\cite{1}.

\section*{Application of the modified RF to the exterior Marder space:}
We try to study the evolution of area of a cylinder of fixed height $z$ and radius $r$ with the parameter $\tau$ using the modified RF equation\cite{1} in exterior Marder space. This is an extension of the work of Samuel and Roy Chowdhury \cite{4}. The difference is that in the present case we extend the formalism for an asymptotically non-flat space.
The curved surface area is
\begin{equation} \int dA= \int_0^{2\pi} \int_0^z \sqrt{g_{zz}g_{\phi \phi}}dz d\phi = 2\pi Bzr^{(c-1)^2}. \end{equation}
The area of each of the caps is
\begin{equation} \int dA= \int_0^{2\pi} \int_0^r \sqrt{g_{rr}g_{\phi \phi}}dr d\phi = \frac{2\pi Bzr^{c^2-2c+2}}{(c-1)^2}. \end{equation}
Therefore, the total area is
\begin{equation} A=2\pi B r^{(c-1)^2}\left(z+\frac{2r}{(c-1)^2}\right). \end{equation}
The compactness C of the surface is given by 
\begin{equation} C=\int_{surface}(2R-K^2) \end{equation}
where, $R$ is the curvature scalar, and $K$ is the trace of the extrinsic curvature.\\
For the cylindrical surface under consideration, $R$ = $0$. $K$ = $D_a \hat n^a$, where $\hat n$ is the unit normal to the surface we are considering. For the curved surface of the cylinder,
\begin{equation} \hat n = \left(\frac{1}{Br^{c(c-1)}},0,0\right), \end{equation}
\begin{equation} K = \frac{(c-1)^2}{Br^{c^2-c+1}}. \end{equation}
For the top surface, the unit normal is
\begin{equation} 
\hat n=\left(0,0,\frac{1}{Br^{c(c-1)}}\right),
\end{equation}
\begin{equation} 
K=0. 
\end{equation}
Therefore,
\begin{equation} C=-\frac{(c-1)^4 2\pi z}{B r^{c^2+1}}. \end{equation}
To calculate $\frac{dA}{d\tau}$, we need to use the Ricci flow equations (4)with the metric (9). For $\mu=\nu=r$, we get
\begin{equation} \frac{dr}{d\tau}=\frac{(c-1)}{B^2}\frac{1}{r^{2c^2-2c+1}}, \end{equation}
whereas, $\mu=\nu=z$ or $\phi$ gives us
\begin{equation}
\frac{dr}{d\tau}=-\frac{c}{B^2}\frac{1}{r^{2c^2-2c+1}}. 
\end{equation}

In order to make equations (23) and (24) consistent, it is easy to see that  $c=\frac{1}{2}$ . Using (23) in (16) one obtains
\begin{equation} 
\frac{dA}{d\tau}=\frac{c}{(c-1)^2}C-\frac{4\pi c}{B}\left(1+\frac{1}{(c-1)^2}\right). 
\end{equation}
For $c$ = $\frac{1}{2}$, we can clearly see that the following inequality holds
\begin{equation} 
\frac{dA}{d\tau} \leq 2C. 
\end{equation}
As $C$ is negative (equation (22)), the area under the modified Ricci flow is a decreasing function of $\tau$.

\section*{Conclusion:}
We started with the modified Ricci flow \cite{1}, and proved that a scalar diffusivity function exists for asymptotically non-flat spaces by finding such a solution for the exterior Marder space. We also used the Ricci flow technique to study the evolution of a cylindrical surface with fixed height, and found an upper bound to the rate of change of area of the surface in terms of the compactness. However, the Ricci flow could be related to the existence of a fixed point solution only for $c=\frac{1}{2}$, which is the limiting case leading to the Raychaudhuri-Som distribution \cite{5}. It deserves mention that for the Marder space, $\frac{1}{2}c$ indicates the mass per unit length of the cylinder, and for $c>\frac{1}{2}$ a photon cannot escape to infinity\cite{11} but rather has a turning point at a finite $r$. This indicates that $c=\frac{1}{2}$ is the limiting case that the cylindrical mass distribution indeed has an event horizon.\\
At this stage one is not sure if an entropy could be defined in the case of a Marder space, but the modified Ricci flow equation 
indeed has a fixed point, and the scalar diffusivity has the significance of the lapse function for the spacetime. This indicates 
that such flow equations need attention in the asymptotically non-flat spaces as well.

\section*{Acknowledgement:}
One of the authors (SC) gratefully acknowledges the warm hospitality provided by IISER-Kolkata, where a major part of the work was done under the Summer Research Programme. We thank the anonymous referee for the suggestions on the earlier draft which indeed resulted in an improvement of the paper.


\begin{thebibliography}{50}
\bibitem{6} P. Figueras, J. Lucietti and T. Wiseman; Class. Quantum Grav. \textbf{28}215018, 2011.
            E. Bahaud; arXiv:1011.2999.
            S. Anastassiou and I. Chrysikos; J. Geom. Phys. \textbf{61}, 1587, 2011.
            W. Graf; PMC \textbf{A1}, 3, 2007.
\bibitem{7} G. Holzegel, T. Schmelzer and C. Warnick; arXiv:0706.1694
\bibitem{8} S. Dutta and V. Suneeta; Class. Quantum Grav., \textbf{27}, 075012, 2010.
            M. Headrich and T. Wiseman; Class. Quantum Grav., \textbf{23}, 6683, 2006.
\bibitem{9} S. Hu, Zhi Hu and R. Zhang; Int. J. Mod. Phys. \textbf{A25}, 2535, 2010.
\bibitem{10}S. Das, K. Prabhu and S. Kar; Int. J Geom. Methods in Mod. Phys. \textbf{7}, 837, 2010.
\bibitem{2} G. Perelman;  Preprint math.DG/0211159, 2002.    
\bibitem{1} J. Samuel  and S. Roy Chowdhury; Class. Quantum Grav., \textbf{24}, F47, 2007. [arXiv:0711.0428].
\bibitem{3} L. Marder;  Proc.R.Soc. A, \textbf{244} 524, 1958.
\bibitem{4} J. Samuel and S. Roy Chowdhury; Class. Quantum Grav., \textbf{25}, 035012, 2008.[arXiv:0711.0430].
\bibitem{5} A. K. Raychaudhuri and M. Som; Proc.Camb.Phil.Soc. \textbf{58} 338, 1962.
\bibitem{11} A. Banerjee; J.Phys.A, \textbf{1}, 495, 1968.
\end{thebibliography}
\end{document}